\documentclass[english,aps,prd,reprint,showpacs,floatfix,nofootinbib]{revtex4-1}

\usepackage{babel}
\usepackage{amsmath}
\usepackage{amssymb}
\usepackage{graphicx}
\usepackage{esint}
\usepackage{longtable}
\usepackage{dcolumn}


\begin{document}

\title{Rotating black holes in a Randall-Sundrum brane with a cosmological constant}

\author{J. C. S. Neves}
\email{nevesjcs@if.usp.br}
\affiliation{Instituto de F\'{\i}sica, Universidade de S\~{a}o Paulo \\
 C.P. 66318, CEP 05315-970, S\~{a}o Paulo-SP, Brazil}

\author{C. Molina}
\email{cmolina@usp.br}
\affiliation{Escola de Artes, Ci\^{e}ncias e Humanidades, Universidade de S\~{a}o Paulo\\
 Av. Arlindo Bettio 1000, CEP 03828-000, S\~{a}o Paulo-SP, Brazil}

\begin{abstract}
In this work we have constructed axially symmetric vacuum solutions of the gravitational field equations in a Randall-Sundrum brane. A non-null effective cosmological constant is considered, and asymptotically de Sitter and anti-de Sitter spacetimes are obtained. The solutions describe rotating black holes in a four-dimensional brane. Optical features of the solutions are treated, emphasizing the rotation of the polarization vector along null congruences.
\end{abstract}


\maketitle

\section{Introduction}
\label{introduction}

Black holes are important objects in the theory of general
relativity and in theories of modified gravity. Recently, strong observational indications of black
holes have been pointed out, for instance, in the centers of galaxies \cite{Kormendy}.
Among the black hole types, those with axial symmetry are considered
more realistic. The most important solution with this symmetry is the Kerr metric, published in 1963
\cite{Kerr}. The Kerr geometry is asymptotically flat, and its
anti-de Sitter (AdS) and de Sitter (dS) generalizations
were made in 1968 by Carter \cite{Carter1}. The latter case, the dS
solution, is more appropriate, for instance, to model a black hole in an expanding universe. Observations published since 1998 \cite{Supernova} suggest that our Universe is dominated by an accelerated expansion, described in the $\Lambda$CDM model by effect of a positive cosmological constant. On the other hand, AdS solutions have been used
in the so-called AdS correspondence \cite{Hawking}. In this
work, we are interested in axially symmetric solutions in a related context: the brane world scenario.

Brane world models have been studied since the 1990s
(see the reviews \cite{Maartens,Clifton}, for instance). In the model presented by Arkani-Hamed, Dimopoulos, and Dvali (ADD) \cite{ADD}, and in several approaches proposed later, 
the brane---our Universe---is embedded in a $(4+d)$-spacetime
called bulk. The $d$ extra dimensions are compact, and all fields
are confined on the brane. In preliminary calculations, the authors
have stressed that for $d>1$ the extra dimensions have radii in
the millimeter scale. One basic motivation for the ADD model was to solve the hierarchy
problem, which is the difference between the electroweak scale ($\sim$ TeV)
and the 4-dimensional Planck scale \linebreak ($\sim 10^{16}$ TeV). According
to this model, the electroweak scale is the fundamental scale, and
the $(4+d)$-dimensional Planck scale has the same order. Another option
to solve the hierarchy problem was the so-called Randall-Sundrum brane model. The ADD picture evolved to scenarios where the extra
dimension can be unbounded.

In the first Randall-Sundrum model (RS-I) \cite{Randall-Sundrum},
 the hierarchy problem was considered with the introduction of a warp factor
in the bulk metric. In this case, the bulk is described
by a five-dimensional AdS spacetime, which has two 3-branes as boundaries
(our Universe and a hidden universe). The extra dimension is compact
and it has finite radius. Like the ADD model, in the RS-I the five-dimensional
Planck scale is the electroweak scale. On other hand, the second Randall-Sundrum model (RS-II) \cite{Randall-Sundrum 2}
adopts only a single 3-brane; in this case the extra dimension radius
is infinity.

One way to construct compact solutions, black holes and wormholes for instance,  in a Randall-Sundrum
scenario, is to consider the four-dimensional effective gravitational field equations deduced
by Shiromizu \textit{et al.} \cite{Shiromizu}. These induced equations were
obtained from Gauss-Codazzi equations in five dimensions. Assuming
a brane world type model (the RS-II, for instance), the authors enforce
a $\mathbb{Z}_{2}$ symmetry in the bulk and obtain the 3-brane
gravitational field equations, which involve brane quantities and
bulk quantities. The complete solution (bulk plus brane) remains unknown
in this approach. Still, Campbell-Magaard theorems \cite{Campbell,Seahra2}
guarantee the extensions of the brane solutions through the bulk,
locally at least.

In this context, spherically symmetric solutions were
studied. In Refs.~\cite{Casadio,Bronnikov}, black
hole and wormhole solutions were constructed, assuming a zero value for the cosmological constant in
the brane. In Ref.~\cite{Molina_Neves}, these solutions were generalized considering a negative value to the cosmological constant in the
brane, and asymptotically anti-de Sitter black holes and wormholes were found. In the same context, wormhole solutions in
an asymptotically de Sitter brane were obtained in Ref.~\cite{Molina_Neves_2}. 
Other constraints of interest were considered in Refs.~\cite{Kaloper, Lemos-2003, Lobo2, Molina3, Prado}.

\newpage

Solutions with axial symmetry were derived in Ref.~\cite{Dadhich_Maartens},
where induced equations by Shiromizu \textit{et al.} were used. In Ref.~\cite{Aliev},
Aliev and G\"{u}mr\"{u}k\c{c}\"{u}oglu, from other effective field equations \cite{Aliev 2}, obtained an asymptotically flat metric with axial symmetry using
the Kerr-Schild ansatz. Assuming a 3-brane vacuum, these authors have
shown the presence of an induced Colombian type charge in the four-dimensional metric. 

This charge---a tidal charge---can be interpreted as the bulk's influence
on the brane. 

Exploring the spacetimes obtained here, we have studied optical features
of the solutions constructed. Following Pineault and Roeder's
work \cite{Pineault}, we have used the Newman-Penrose formalism \cite{NP}
adapted to the  locally nonrotating frames \cite{Bardeen} to obtain the rotation
of the polarization vector of the light in the geometrical optics regime.
A geometrical optics description is appropriate for high frequency
electromagnetic waves. In this approach, the light propagates along
null geodesics and its polarization vector is parallelly transported.
Once again, the results show that the bulk's influence on the brane is described  by a tidal charge. 

The structure of this paper is presented in the following. In Sec.~\ref{solution} we derive asymptotically de Sitter and anti-de Sitter solutions
with axial symmetry which satisfy specified conditions on the brane. In Sec.~\ref{optics}, the presented spacetimes are explored, with applications involving geometrical optics. In Sec.~\ref{remarks} final comments are made. In Appendix \ref{geodesics} the geodesics equations, used in our work, are indicated; and in
Appendix \ref{non-rotating}, important quantities for the calculations in the locally nonrotating frame are showed.

In this work we have used the metric signature $\textrm{diag} \, (-+++)$ and the
geometric units $G_{4D}=c=1$, where $G_{4D}$ is the effective four-dimensional
gravitational constant. Greek indices are tensor indices (run from
0 to 3); Latin indices indicate tetrad components (run from 1 to 4).

\section{Rotating solutions on the brane}
\label{solution}

The effective four-dimensional gravitational field equations projected
on the brane, considering a Randall-Sundrum type model, were deduced
by Shiromizu \textit{et al.} in \cite{Shiromizu}, where one brane
and a five-dimensional AdS bulk with $\mathbb{Z}_{2}$ symmetry were
used. For these authors, the induced vacuum field equations on the brane are 
\begin{equation}
G_{\mu\nu} = -\Lambda_{4D}g_{\mu\nu} - E_{\mu\nu} \,\, ,
\label{eq_projetada}
\end{equation}
where $G_{\mu\nu}$ is the four-dimensional Einstein tensor associated
with the brane metric $g_{\mu\nu}$, $\Lambda_{4D}$ is the four-dimensional
brane cosmological constant, and $E_{\mu\nu}$ is proportional to
the (traceless) projection on the brane of the five-dimensional Weyl
tensor. The vacuum field equations (\ref{eq_projetada}) reduce
to vacuum Einstein equations in the low energy limit. 

A combination of the effective Einstein equations (\ref{eq_projetada}) written without specifying $E_{\mu\nu}$ is the  trace of Eq.~(\ref{eq_projetada}):
\begin{equation}
R=4\Lambda_{4D} \,\, .
\label{Ricci}
\end{equation}
To solve Eq.~(\ref{Ricci}) in both dS and AdS scenarios, we assume axial symmetry in the
brane. Following \cite{Aliev}, we use the Kerr-Schild-(A)-dS ansatz \cite{Gibbons}
\begin{equation}
ds^{2} = ds_{\Lambda}^{2} + H \left( l_{\mu}dx^{\mu} \right)^{2} \,\, ,
\label{Kerr-Schild}
\end{equation}
where $ds_{\Lambda}^{2}$ indicates the AdS or dS metric, $H$ is a function
of $r$ and $\theta$, and $l_{\mu}$ represents a null vector field. In
the $(\tau,r,\theta,\phi)$ coordinate system, Eq.~(\ref{Kerr-Schild})
can be written as
\begin{eqnarray}
ds_{\Lambda}^{2} & = & -\frac{\left(1-\frac{\Lambda_{4D}}{3}r^{2}\right)\Delta_{\theta}}{\Xi} d\tau^{2} 
+ \frac{\Sigma}{\left(1-\frac{\Lambda_{4D}}{3}r^{2}\right) \left(r^{2}+a^{2}\right)} dr^{2} \nonumber \\
& & + \frac{\Sigma}{\Delta_{\theta}} d\theta^{2} + \frac{\left(r^{2} + a^{2}\right) \sin^{2} \theta}{\Xi} d\phi^{2}
\label{ds_linha}
\end{eqnarray}
and
\begin{eqnarray}
H(l_{\mu}dx^{\mu})^{2} & = & H\left[ \frac{\Delta_{\theta}}{\Xi} d\tau + \frac{\Sigma}{ \left(1-\frac{\Lambda_{4D}}{3}r^{2} \right) (r^{2} + a^{2})}dr \right.
\nonumber \\
& & \left. - \frac{a \sin^{2}\theta}{\Xi} d\phi \right]^{2} \,\, ,
\label{H-1}
\end{eqnarray}
where
\begin{gather}
\Delta_{\theta} = 1 + \frac{\Lambda_{4D}}{3} a^{2} \cos^{2} \theta \,\, ,
\hspace{0.5cm}
\Sigma = r^{2} + a^{2} \cos^{2} \theta \,\, ,
\nonumber \\
\Xi = 1 + \frac{\Lambda_{4D}}{3} a^{2} \,\, .
\label{definitions}
\end{gather}

Substituting Eqs.~(\ref{ds_linha})-(\ref{H-1}) into Eq.~(\ref{Kerr-Schild}),
we have a partial differential equation for the $H$ function, that
is, 
\begin{equation}
R = \frac{\partial^{2}H}{\partial r^{2}} + \frac{4r}{\Sigma} \frac{\partial H}{\partial r} + \frac{2}{\Sigma}H + 4\Lambda_{4D} \,\, .
\label{equation1}
\end{equation}
From Eq.~(\ref{Ricci}), Eq.~(\ref{equation1}) can be written
as the homogeneous equation
\begin{equation}
\frac{\partial^{2}H}{\partial r^{2}} + \frac{4r}{\Sigma}\frac{\partial H}{\partial r} + \frac{2}{\Sigma} H = 0 \,\,.
\label{equation2}
\end{equation}
This equation has the following general solution:
\begin{equation}
H = \frac{2Mr}{\Sigma} - \frac{q}{\Sigma} \,\, ,
\label{H}
\end{equation}
where the parameters $M$ and $q$ are constants of integration. Physical interpretations for these constants will be introduced with the Boyer-Lindquist coordinates. The expression in Eq.~(\ref{H}) has the same form of the analogous null $\Lambda_{4D}$ result, obtained in \cite{Aliev}. 

The tensor $E_{\mu\nu}$, which can be interpreted as an effective stress energy tensor in the context of the general relativity with the usual Einstein equations, satisfies by construction the condition  $E_{\mu}^{\mu}=0$. Its explicit form is given by
\begin{eqnarray}
E_{t}^{t} & = & -E_{\varphi}^{\varphi} = q \left[ \frac{2(r^{2} + a^{2})}{\Sigma^{3}} - \frac{1}{\Sigma^{2}} \right] \,\, , \\
E_{r}^{r} & = & -E_{\theta}^{\theta} = q \left[ \frac{2(r^{2} + a^{2})}{\Sigma^{3}} -\frac{1}{\Sigma^{2}} \right] \,\, , \\
E_{\varphi}^{t} & = & -\frac{\left(r^{2} + a^{2}\right) \sin^{2}\theta}{\Xi} \,\, , \\
E_{t}^{\varphi} & = & -\frac{2qa}{\Xi\Sigma^{3}}\left(r^{2} + a^{2}\right) \sin^{2}\theta \,\, .
\label{Componentes E}
\end{eqnarray}
It should be stressed that, in a brane world context, we assume vacuum in the four-dimensional hypersurface. The quantity $E_{\mu\nu}$ is understood as a geometric component
in Eq.~(\ref{eq_projetada}); it is the part of five-dimensional Weyl
tensor projected on the brane. In this setting, $E_{\mu\nu}$ expresses the bulk influence on
the brane.

In the Boyer-Lindquist coordinates the structure of the
spacetime is seen more clearly. Event and Killing
horizons are immediately pointed out with this chart. 
After making the transformations
\begin{equation}
d\tau = dt + \frac{\Sigma H}{\left(1 - \frac{\Lambda_{4D}}{3}r^{2}\right) \Delta_{r}} dr \,\, ,
\label{transf_1}
\end{equation}
\begin{equation}
d\phi = d\varphi - \frac{\Lambda_{4D}}{3}adt + \frac{a\Sigma H}{(r^{2}+a^{2}) \Delta_{r}} dr \,\, ,
\label{transf_2}
\end{equation}
where we define
\begin{equation}
\Delta_{r} = (r^{2} + a^{2})\left( 1 - \frac{\Lambda_{4D}}{3}r^{2}\right) - 2Mr + q \,\, ,
\label{Delta_r}
\end{equation}
the metric (\ref{Kerr-Schild}) takes the familiar form
\begin{widetext}
\begin{eqnarray}
ds^{2} & = & -\frac{1}{\Sigma}\left(\Delta_{r} - \Delta_{\theta} a^{2} \sin^{2} \theta\right) dt^{2} - \frac{2a}{\Xi\Sigma} \left[(r^{2} + a^{2})\Delta_{\theta} - \Delta_{r}\right] \sin^{2} \theta dtd \varphi
+ \frac{\Sigma}{\Delta_{r}}dr^{2} + \frac{\Sigma}{\Delta_{\theta}}d\theta^{2}
\nonumber \\
& &  + \frac{1}{\Xi^{2}\Sigma}\left[(r^{2}+a^{2})^{2}\Delta_{\theta}-\Delta_{r}a^{2}\sin^{2}\theta\right]\sin^{2}\theta d\varphi^{2} \,\, 
.\label{Metrica_Boyer-Lindquist}
\end{eqnarray}
\end{widetext}
This metric has the same form of the the Kerr-Newman-(A)-dS metric in the general
relativity context. In the Boyer-Lindquist coordinates the constant $q$ 
plays the role of a Colombian charge. However, there are neither electric
or magnetic fields on the brane. The field equations in (\ref{eq_projetada})
assume a vanishing stress energy tensor on the brane. In this sense,
the origin of $q$ is the five-dimensional spacetime, the bulk. 
The constant $q$ is an induced charge on the brane, a tidal charge, encoding the bulk's influence on the brane.
In this brane world context, $q$ can assume either positive or negative values. As will be seen in the following, the
negative $q$ case amplifies the gravitational effects on the brane.
The constant $M$, in the asymptotically flat solution, is the conserved
quantity recognized as mass (see for example \cite{Caldarelli,Dehghani}).

We observe that the metric functions in Eq.~(\ref{Metrica_Boyer-Lindquist}) diverge as $\Delta_{r} \rightarrow 0$, $\Delta_{\theta} \rightarrow 0$, and $\Xi \rightarrow 0$. The behavior of the Kretschmann scalar in this limit shows that there is a curvature singularity with $r \rightarrow 0$ and $\cos \theta \rightarrow 0$. That is,
\begin{equation}
R_{\alpha\beta\gamma\delta} R^{\alpha\beta\gamma\delta} \propto 
\frac{1}{\Sigma^{8}} + \frac{1}{\Delta_{\theta}} \,\, .
\label{Kretschmann}
\end{equation}
The condition $\Delta_{\theta}=0$ is not feasible in the dS case $(\Lambda_{4D}>0)$. The expression in (\ref{Kretschmann}) indicates that the divergence of 
 the Kretschmann scalar with $\Delta_{r} \rightarrow 0$
is a coordinate system problem. In a convenient chart, such as the
Kerr-Schild-(A)-dS for instance, the metric is regular. The spacetime equipped with the metric in 
(\ref{Metrica_Boyer-Lindquist}) presents a ringlike singularity \cite{Hawking_Ellis}.

Also from Eq.~(\ref{Metrica_Boyer-Lindquist}), we can find out the real positive
roots of $g^{rr}$, which provide the horizon radii. That is, from
Eq.~(\ref{Delta_r}), the relevant condition is $\Delta_{r}=0$. The roots structure of the function $\Delta_{r}$ depends strongly on the sign of the effective cosmological constant. We will treat the two cases,  $\Lambda_{4D}<0$ and $\Lambda_{4D}>0$, in the following.

\subsection{Asymptotically anti-de Sitter geometry}
\label{AdS_geometry}

When $\Lambda_{4D}<0$, that is, considering asymptotically anti-de Sitter spacetimes, and $a<M$,  the function $\Delta_{r}$ has two real roots and two complex roots, since $q_{min}<q<q_{max}$ (the values of $q_{min}$ and
$q_{max}$ depend on $a,M,$ and $\Lambda_{4D}$). The real
roots $(r_{-},r_{+})$ are associated with an internal Cauchy horizon and an external event horizon, respectively. The constant $q$  is determinant on the location
of these horizons, as illustrated in Fig.~1. Negative values of $q$ render roots larger than positive values and increase the bulk's influence on the brane, a point commented on also in \cite{Aliev}.
The bulk's effects---geometrical effects---are
amplified in this case. In this case, the spacetime structure seems like the Kerr-Newman-AdS structure. However, when $q<q_{min}$, we have something different: $\Delta_{r}$ has only one root, according to Fig.~1. 

\begin{figure}[ht]
\begin{center}
\includegraphics[clip,width=\columnwidth]{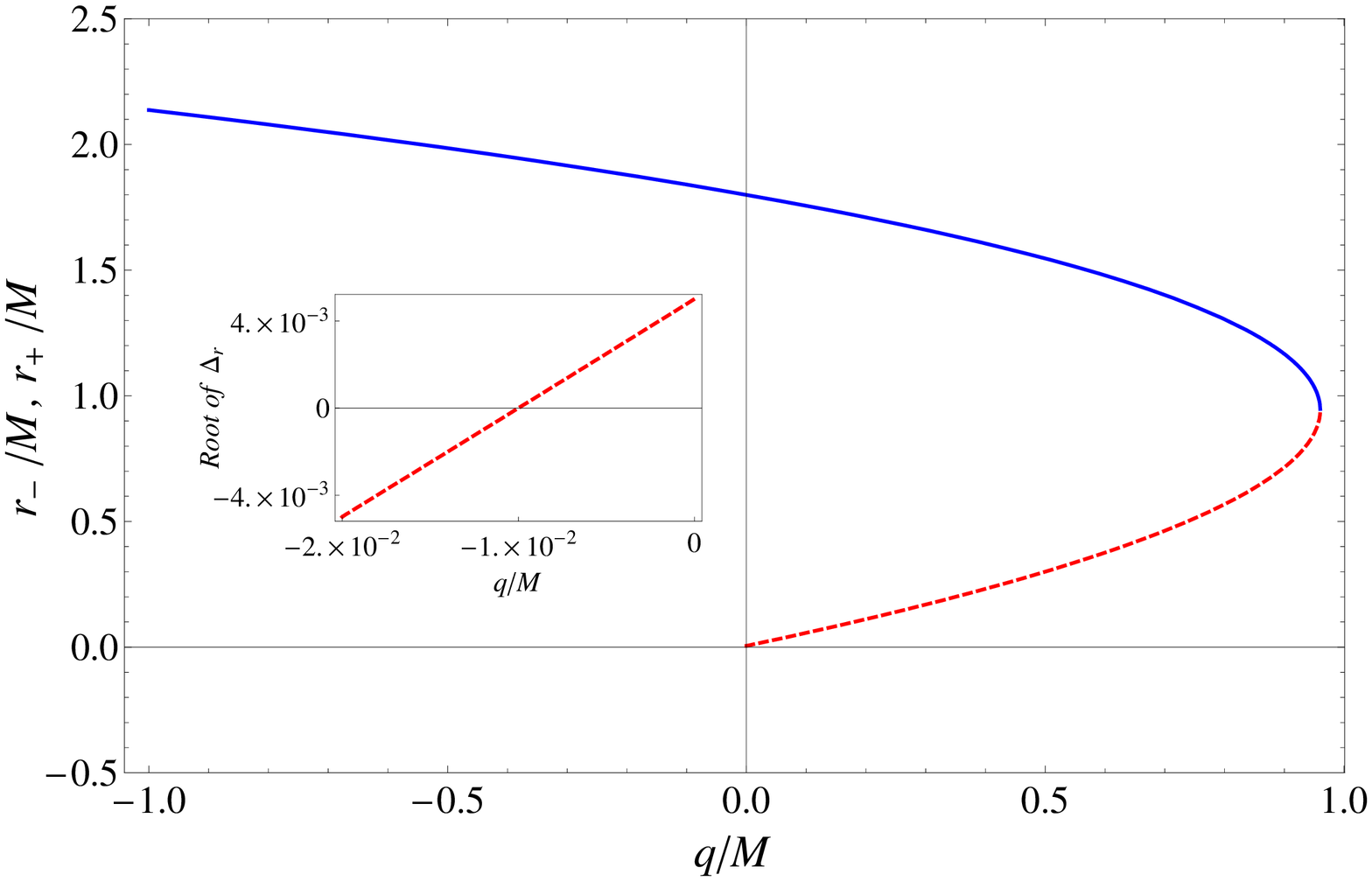} 
\end{center}
\caption{Typical dependence of the horizon radii, $r_{-}$ (Cauchy horizon) and $r_{+}$ (event horizon), with the parameter $q$ when $\Lambda_{4D}<0$. In the detail, the existence of a positive $r_{-}$ is indicated. In this plot, $a=M/10$, $\Lambda_{4D}=-M/10$ and \mbox{$\theta=\pi/2$}.}
\end{figure}

The metric in (\ref{Metrica_Boyer-Lindquist}) is independent of $t$
and $\varphi$, a result reminiscent of the existence of two Killing vector fields $\xi_{t}$ and $\xi_{\varphi}$.  In coordinate basis, $\xi_{t} = \partial/\partial t$ and \linebreak $\xi_{\varphi} = \partial/\partial\varphi$. The vector field  $\xi_{\varphi}$ is spacelike, defining the axial symmetry. The norms of those
vector are
\begin{eqnarray}
\xi_{t}^{2} = g_{tt} & = & -\frac{1}{\Sigma} \left( \Delta_{r} - \Delta_{\theta} a^{2}\sin^{2}\theta \right) \,\, ,
\label{Killing_norm1}\\
\xi_{\varphi}^{2} = g_{\varphi\varphi} & = & \frac{1}{\left( 1 + \frac{\Lambda_{4D}}{3}a^{2} \right)^{2} \Sigma} \left[(r^{2} + a^{2})^{2} \Delta_{\theta} 
\right.
\nonumber \\
& & \left.
- a^{2}\Delta_{r}\sin^{2}\theta\right] \sin^{2} \theta \,\, .
\label{Killing_norm2}
\end{eqnarray}
A Killing horizon is the surface where the tangent Killing vector
field is lightlike. Setting $\xi_{t}^{2}$ and $\xi_{\varphi}^{2}$ in Eqs.~(\ref{Killing_norm1}) and (\ref{Killing_norm2})
to zero, we obtain the localization of the Killing horizons. For
the Killing vector $\xi_{t}$, the roots of $\xi_{t}^{2}$
depend on $a$, $M$, $\theta$, $\Lambda_{4D}$, and $q$. For values of $q/M$
which allow us to have two roots for $\xi_{t}^{2}$, the spacetime is composed of five regions, divided by $r$-constant surfaces $r=S_{+}$, $r=r_{+}$, $r=r_{-}$ and $r=S_{-}$ with
\begin{equation}
0 < S_{-} < r_{-} < r_{+} < S_{+} < \infty \,\, .
\label{AdS_region}
\end{equation}
The surfaces $r=S_{-}$ and $r=S_{+}$, in the maximal extension of the considered geometries, are Killing horizons. The important
region between $r=r_{+}$ and $r=S_{+}$ is the ergosphere. In this region, the field $\xi_{t}$ becomes spacelike. In the case when $q<q_{min}$ there is only one root for (\ref{Killing_norm1}), and the spacetime structure is given by
\begin{equation}
0<r_{+}<S_{+}<\infty.
\label{AdS_region2}
\end{equation}
Once again, the ergosphere is the region between $r=r_{+}$ and $r=S_{+}$.

\subsection{Asymptotically de Sitter geometry}
\label{dS_geometry}

With $\Lambda_{4D}>0$, that is, considering asymptotically de Sitter spacetimes, we have at most three positive roots for the function $\Delta_{r}$. Focusing on $a<M$, when $q_{min}<q<q_{max}$ (where the values of $q_{min}$ and $q_{max}$
depend on $a$, $M$, and $\Lambda_{4D}$), the function $\Delta_{r}$
indicates the existence of an internal horizon ($r=r_{-}$), an event horizon ($r=r_{+}$),
and a cosmological horizon ($r=r_{c}$). Once again, negative values for $q$ allow larger horizon
radii. Another important feature is the relative size between
$r_{+}$ and $r_{-}$. Such a larger event horizon means a smaller cosmological
horizon. This is a property from the Girard-Newton-Vi\'{e}te
polynomial formulas. On the other hand, when $q_{lim2}<q<q_{lim1}$, where $q_{lim1}$ and $q_{lim2}$ depend on $a,M$, and $\Lambda_{4D}$, the function $\Delta_{r}$ has at most two real roots, $r_{+}$ and $r_{c}$. In the latter case, as we will see, the ergosphere is absent. Finally, when $q_{lim1}<q<q_{min}$, $\Delta_{r}$ has two real roots, and there is an ergosphere in this interval of parameters, as we will see in the following.

\begin{figure}[ht]
\begin{center}
\includegraphics[clip,width=\columnwidth]{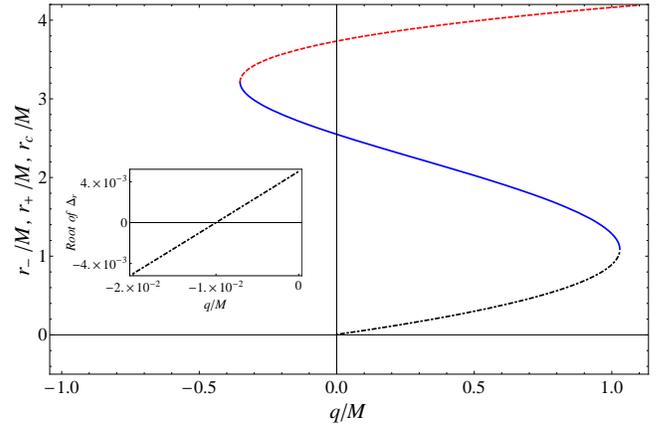}
\end{center}
\caption{Typical dependence of the horizon radii, $r_{-}$ (Cauchy horizon), $r_{+}$ (event horizon) and $r_{c}$ (cosmological horizon), with the parameter $q$ when $\Lambda_{4D}>0$. In the detail, the existence of a positive $r_{-}$ is indicated. In this plot, $a=M/10$, $\Lambda_{4D}=M/10$ and $\theta=\pi/2$.}
\end{figure}

In the asymptotically de Sitter geometry, there are at most three positive roots of $\xi_{t}^{2}$, which depend on $a$, $M$, $\theta$, $\Lambda_{4D}$ and $q$. Thus, we can divide the spacetime in six regions, separated by $r$-constant surfaces $r=r_{c}$, $r=S_{+}$, $r=S_{i}$, $r=r_{+}$, $r=r_{-}$, $r=S_{-}$ with
\begin{equation}
0<S_{-} < r_{-} < r_{+} < S_{i} < S_{+} < r_{c} \,\, .
\label{dS_region}
\end{equation}
The $r=S_{-}$, $r=S_{i}$ and $r=S_{+}$ surfaces are Killing horizons from
the Killing vector field $\xi_{t}$. There is an important region between
$S_{i}$ and $S_{+}$, where the Killing vector
$\xi_{t}$ becomes spacelike. Thus, considering Eq.~(\ref{dS_region}), the ergosphere
doesn't reach the event horizon. 

On the other hand, when $q_{lim2}<q<q_{lim1}$, the Killing vector field $\xi_{t}$ is timelike everywhere. Thus, there is no ergosphere for these values of $q$. However, when $q_{lim1}<q<q_{min}$, the function $\xi_{t}^{2}$ has two positive roots. The spacetime structure in this case is characterized by
\begin{equation}
0 <r_{+}< S_{-}< S_{+}< r_{c} \,\, . 
\label{dS_region2}
\end{equation}  
The Killing vector field $\xi_{t}$ is timelike in $0<r<S_{-}$ and $S_{-}<r<r_{c}$; and it is spacelike in the $S_{-}<r<S_{+}$ region. The ergosphere does not reach the event horizon.

\section{Geometrical optics: rotation of polarization vector}
\label{optics}

In order to analyze possible effects of the extra dimensions in terms of measurable quantities, we focus on geometrical optics. The approach employed here was presented by Pineault and Roeder in \cite{Pineault}, where the Newman-Penrose formalism was used to calculate optical quantities in the weak-field approximation with $a \ll M$. The equations which govern the tangent vector $k^{\mu}$ (the wave vector) to the null congruence and the polarization vector $f^{\mu}$ are
\begin{equation}
k^{\mu}k_{\mu} = 0 \,\, , \,\,
Dk^{\mu} = 0
\label{OG1}
\end{equation}
and
\begin{equation}
k^{\mu} f_{\mu} = 0 \,\, , \,\,
Df^{\mu}=0 \,\, ,
\label{OG2}
\end{equation}
with the operator $D$ denoting covariant derivative in the $k^{\mu}$
direction. 

In the Newman-Penrose formalism \cite{NP} a null tetrad
is adopted, $\left\{ e_{a\mu}\right\} =\left(m_{\mu},\bar{m}_{\mu},l_{\mu},k_{\mu}\right)$, with the vector $m_{\mu}$ given by
\begin{equation}
m^{\mu}=\frac{\sqrt{2}}{2}(a^{\mu}+ib^{\mu}) \,\, .
\label{def_m}
\end{equation}
The vector $m_{\mu}$ is particularly relevant to the work developed here, as will be seen. An important feature of the formalism is that the $k^{\mu}$ direction is preserved
under null rotations as
\begin{gather}
k'^{\mu}=Ak^{\mu} \,\, , \hspace{0.5cm}
m'^{\mu}=e^{-i\chi}\left(m^{\mu}+Bk^{\mu}\right) \,\, , 
\nonumber \\
l'^{\mu} = A^{-1}\left(l^{\mu}+B\bar{m}^{\mu}+\bar{B}m^{\mu}+B\bar{B}k^{\mu}\right) \, \, ,
\label{null_rotation}
\end{gather}
with $A>0$, $B$ complex, and $\chi$ real. The Newman-Penrose formalism
provides 12 constants, the spin coefficients, to the characterization of the spacetime. Some coefficients will be used to estimate the variation of the polarization vector. 

As shown in \cite{NP}, when $k^{\mu}$ is tangent
to the null congruence the spin coefficient \linebreak $\kappa\equiv-Dk_{\mu}m^{\mu}$
is zero. Moreover, considering that the null tetrad must parallelly
propagate along the null congruence, this assumption implies that
another two spins coefficients vanish: $\epsilon=\pi=0$. Then, the
plane spanned by $k^{\mu}$ and $a^{\mu}$ can be identified with
the polarization plane, which is parallelly propagated in the $k^{\mu}$
direction. That is, the polarization vector can be identified with
the $a^{\mu}$ vector of the Newman-Penrose formalism. From this, we can build an orthonormal frame 
\begin{equation}
\left\{ \boldsymbol{e}_{a}^{\ (\mu)}\right\} = \left\{ r^{(\mu)},\bar{r}^{(\mu)},q^{(\mu)},p^{(\mu)}\right\} \,\, ,
\label{tetrad}
\end{equation}
such that this tetrad corresponds to the one-forms $ \omega^{(0)} = e^{\nu}dt $, $ \omega^{(1)} = e^{\lambda}dr $, $ \omega^{(2)} = e^{\mu}d\theta $, and $ \omega^{(3)} = (d\varphi-\Omega dt)e^{\psi} $ of the locally nonrotating frame (LNRF) \cite{Bardeen}
\footnote{The LNRF indices are indicated between parentheses.}. 
For the metric in Eq.~(\ref{Metrica_Boyer-Lindquist}), the expressions for $e^{\nu}$, $e^{\lambda}$, $e^{\mu}$, and $e^{\psi}$ are presented in the Appendix B.
Therefore, if the source and the observer are at rest with the LNRF, they are rotating with the black hole. In \cite{Pineault} this construction was made, with the expression for the $m_{+}^{(\mu)}$ vector (i.e., $a^{\mu}$)---the projection of $m^{\mu}$ on the LNRF---given by
\begin{eqnarray}
a_{+}^{\ (\mu)} & = & \frac{1}{\sqrt{2}}\left(0,-\frac{k^{(2)}}{k^{(0)}},1-K\left(k^{(2)}\right)^{2},Kk^{(2)}k^{(3)}\right) \,\, , \nonumber \\ \\
b_{+}^{\ (\mu)} & = & \frac{1}{\sqrt{2}}\left(0,-\frac{k^{(3)}}{k^{(0)}},-Kk^{(2)}k^{(3)},1-K\left(k^{(3)}\right)^{2}\right) \,\, , \nonumber \\
\end{eqnarray}
where $K=1/\left[k^{(0)}\left(k^{(0)}+k^{(1)}\right)\right]$, and $k^{(\mu)}$ is the projection of $k^{\mu}$ on the LNRF, according to Eq.~(\ref{k_projetado}). A null rotation was performed
\begin{equation}
m_{+}^{(\mu)}\rightarrow m^{(\mu)} = e^{-i\chi} m_{+}^{(\mu)} \,\, ,
\label{null_rotation 2}
\end{equation}
such that $\epsilon=0$. With this choice, and considering the form
of the $\epsilon$ coefficient \linebreak ($\epsilon\equiv Dm_{\mu}\bar{m}^{\mu}/2$), it is shown that
\begin{equation}
D\chi=-2i\epsilon_{+} \,\, .
\label{D_xi}
\end{equation}

Expression (\ref{D_xi}) indicates how the $\chi$ angle varies
in the $k^{\mu}$ direction, the congruence direction. This variation will be important to calculate
the variation of polarization vector in that direction. Considering the metric presented in the previous section, we obtain
\begin{widetext}
\begin{eqnarray}
D\chi = -2i\epsilon_{+} & = & \frac{\left(\Gamma_{\ \ \ (\theta)(t)}^{(t)}k^{(t)}+\Gamma_{\ \ \ (\theta)(r)}^{(r)}k^{(r)}+\Gamma_{\ \ \ (\theta)(\theta)}^{(r)}k^{(\theta)}+\Gamma_{\ \ \ (\theta)(\varphi)}^{(t)}k^{(\varphi)}\right)k^{(\varphi)}}{k^{(t)}+k^{(r)}}\nonumber \\
 &  & - \frac{\left(\Gamma_{\ \ \ (\varphi)(t)}^{(r)}k^{(t)}+\Gamma_{\ \ \ (\varphi)(r)}^{(t)}k^{(r)}+\Gamma_{\ \ \ (\varphi)(\theta)}^{(t)}k^{(\theta)}+\Gamma_{\ \ \ (\varphi)(\varphi)}^{(r)}k^{(\varphi)}\right)k^{(\theta)}}{k^{(t)}+k^{(r)}}\nonumber \\
 & & + \Gamma_{\ \ \ (\varphi)(\varphi)}^{(\theta)}k^{(\varphi)}+\Gamma_{\ \ \ (\varphi)(t)}^{(\theta)}k^{(t)} \,\, ,
\label{D_xi1}
\end{eqnarray}
\end{widetext}
where the values of $\Gamma_{\ (\nu)(\gamma)}^{(\mu)}$ and $k^{(\mu)}$
were projected on the LNRF, shown in Appendix
\ref{non-rotating}. Using the results in Eqs.~(\ref{k_projetado}) and (\ref{Conec=0000E7=0000F5es}), $D\chi$ is given, in first order in $a$, by 
\begin{eqnarray}
D\chi & = & -\frac{\Phi \cos\theta}{r^{2} \sin^{2} \theta} + \left(3Mr-2q\right)\frac{\sqrt{Q-\Phi^{2} \cot^{2}\theta}}{r^{5}} a \sin \theta 
\nonumber \\
& & + \mathcal{O}(a^2) \,\, .
\label{D_xi2}
\end{eqnarray}
In the regime $a \ll M$, we observe that $\Sigma \approx r^{2}$ and
\begin{equation}
\sqrt{Q - \Phi^{2} \cot^{2} \theta} = r^{2} \dot{\theta} = r^{2}\frac{d\theta}{d\mu} \,\, .
\label{Theta_simpl}
\end{equation}
With the simplification in Eq.~(\ref{Theta_simpl}), we have
\begin{equation}
D\chi = - \frac{\Phi \cos \theta}{r^{2} \sin^{2}\theta} + \left(3M - \frac{2q}{r}\right) \frac{a \sin\theta}{r^{2}}\frac{d\theta}{d\mu} + \mathcal{O}(a^2) \,\, .
\label{D_xi3}
\end{equation}
The expression in Eq.~(\ref{D_xi3}) is associated with the variation between $a^{(\mu)}$
and $a_{+}^{\ (\mu)},$ according to the null rotation indicated in Eq.~(\ref{null_rotation 2}).

The total variation of polarization vector, taking into account Eq.~(\ref{D_xi3}) and the spacetime dragging, is given by
\begin{equation}
\Delta\omega=\Delta\chi+\Delta\varphi \,\, .
\label{Delta}
\end{equation}
The second term in the right side of Eq.~(\ref{Delta}) is due
to the spacetime dragging. The first term is obtained with the integration
of $D\chi$ in Eq.~(\ref{D_xi3}) with the aid of the $\psi$ coordinate (see \cite{Pineault}),
which plays the role of the azimuthal angle in the orbital plane of
the null congruence. Moreover, a new angle was defined: $\alpha$
is the angle of the orbital plane with respect to the equatorial plane.
That is, $\sin \alpha = \cos \theta \sin \psi$, and Eq.~(\ref{D_xi3})
is reduced to
\begin{equation}
D\chi=-\frac{\Phi \cos \theta}{r^{2} \sin^{2} \theta} + D \chi' + \mathcal{O}(a^2) \,\, ,
\label{D_xi4}
\end{equation}
where
\begin{equation}
D\chi'=-\left(3M-\frac{2q}{r}\right)\frac{a \sin \alpha}{r^{2}}\frac{d(\sin \psi)}{d\mu} \,\, .
\label{D_xi5} 
\end{equation}

\begin{figure}[ht]
\begin{center}
\includegraphics[clip,width=0.8\columnwidth]{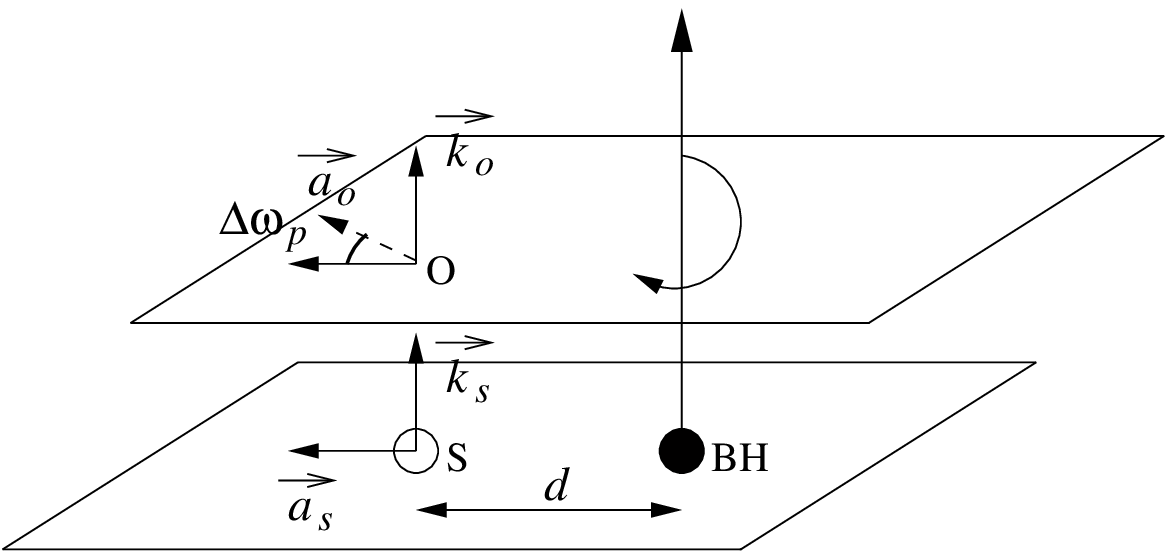}

\vspace{1.5cm}

\includegraphics[clip,width=0.8\columnwidth]{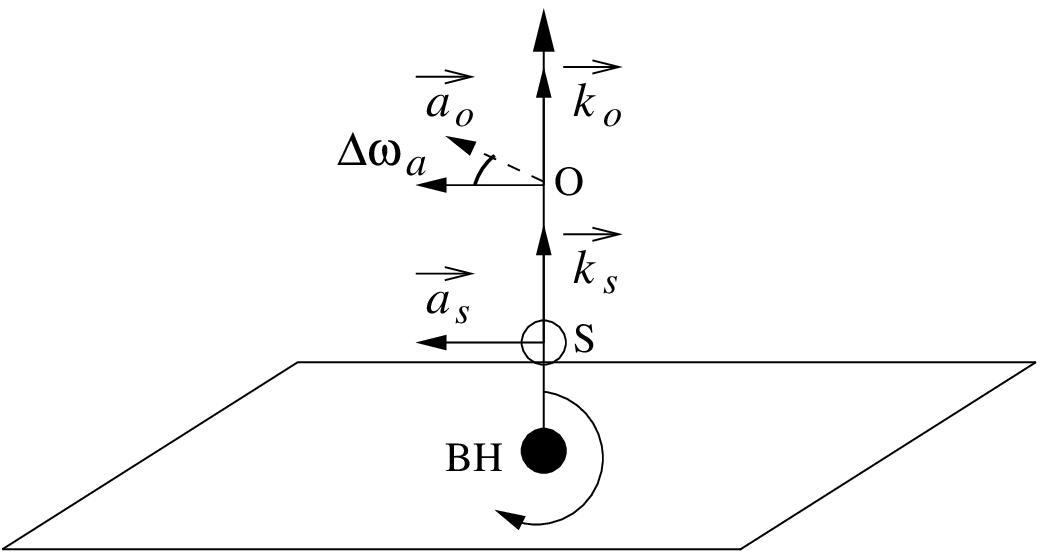}
\end{center}
\caption{At top, the source and the observer are at distance $d$ from
the symmetry axis. At bottom, both source and observer are in the
symmetry axis. The vector tangent to the null congruence and polarization
vector are indicated by $\protect\overrightarrow{k}$ and $\protect\overrightarrow{a}$, respectively. The black hole, the source, and the observer are indicated
by BH, S, and O, respectively.}
\label{scenarios}
\end{figure}

Concretely, we have studied two physical scenarios, illustrated in Fig.~\ref{scenarios}. In the first case, the source is on the
equatorial plane $(\theta=\alpha=\pi/2)$, and the observer is on the plane with $\theta < \pi/2$. Both the source and the observer are at a distance $d$ from the symmetry axis. In the second case, the source and the observer are in the symmetry axis, $r_{o} > r_{s}$. 

The source and the observer are modeled by timelike curves $r=r_{s}$ and $r=r_{o}$, respectively. In both scenarios considered, we assume $a \ll M$ (small rotations), and small values for the cosmological constant $\Lambda_{4D}$. There is no azimuthal component in the photon trajectory, that is, $\Phi = 0$ in the geodesic equations (\ref{t_pt})-(\ref{varphi_pt}). Moreover, we have used the weak-field approximation.

\subsection{Source in the black hole equatorial plane}

We first consider the first proposed scenario, where the source is on the
equatorial plane, and the observer is on the plane with $\theta < \pi/2$ (see Fig.~\ref{scenarios}, top panel). Within the hypothesis assumed, we can write
\begin{equation}
\sin \psi = \frac{\sqrt{r^{2}-d^{2}}}{r} \,\, ,
\label{weak_field}
\end{equation}
which allows us to express $D\chi'$ in Eq.~(\ref{D_xi5}) as
\begin{equation}
D\chi'=-\frac{a}{r^{2}}\left(3M-\frac{2q}{r}\right)\left(\frac{d^{2}}{r^{2}}\frac{1}{\sqrt{r^{2}-d^{2}}}\right)\frac{dr}{d\mu} \,\, ,
\label{D_xi6}
\end{equation}
that is
\begin{equation}
\int_{0}^{\chi}d\chi'=-\int_{d}^{r_{o}}\left(\frac{3Mr-2q}{r^{5}}\right)\frac{ad^{2}}{\sqrt{r^{2}-d^{2}}}\ dr \,\, ,
\label{Integral}
\end{equation}
where $D\chi'=k^{1}d\chi'/dr$ and $k^{1}=dr/d\mu=\dot{r}$ (see Appendix~\ref{geodesics}). Integrating Eq.~(\ref{Integral}), we have the variation between $a^{(\mu)}$
and $a_{+}^{\ (\mu)}$ along the null congruence
\begin{eqnarray}
\Delta\chi & = & -\left\{ \sqrt{r_{o}^{2}-d^{2}}\left[\frac{M}{r_{o}}\left(\frac{1}{r_{o}^{2}}+\frac{2}{d^{2}}\right)-\frac{q}{2r_{o}^{2}}\left(\frac{1}{r_{o}^{2}}+\frac{3}{2d^{2}}\right)\right]\right.\nonumber\\
 & & - \frac{3q}{4d^{3}}\left[\frac{\pi}{2} - \arctan \left(\frac{d}{\sqrt{r_{o}^{2}-d^{2}}}\right)\right]\biggl\} a.
\label{D_xi_total}
\end{eqnarray}
This result, which is $\Lambda_{4D}$ independent, is the same obtained in \cite{Pineault}, since $q=0$, and $r_{o}\rightarrow\infty$ \linebreak ($\Delta\chi=-2Ma/d^2$). Negative values of $q$ amplify the variation between $a^{(\mu)}$ and $a_{+}^{\ (\mu)}$ along the null congruence. 

The effect of dragging, $\Delta\varphi$, is calculated using Eq.~(\ref{k_projetado}), as
%
%\begin{widetext}
\begin{eqnarray}
\Delta\varphi & = &  \int_{\varphi_{i}}^{\varphi_{f}}d\varphi 
%\nonumber \\
%
 = \int_{d}^{r_{o}}\left[\frac{2Mr-q}{r^{3}}+\frac{\Lambda_{4D}}{3}r\right]\frac{aE}{\sqrt{E^{2}r^{2}-d^{2}}}dr
\nonumber \\
 & = & \frac{a}{E} \left\{ \left[\left(\frac{2M}{d^{2}}
-\frac{q}{2d^{2}r_{o}}\right)E^{2} 
+ \frac{\Lambda_{4D}}{3}r_{o}\right]\sqrt{E^{2} - \frac{d^{2}}{r_{o}^{2}}}
\right.
\nonumber \\
& & - \biggr[\left(\frac{2M}{d^{2}}-\frac{q}{2d^{3}}\right)E^{2}
+ \frac{\Lambda_{4D}}{3}d\biggr]\sqrt{E^{2}-1}
\nonumber \\
& & -\frac{qE^{4}}{2d^{3}}\biggl[\arctan \left(\frac{1}{\sqrt{E^{2}-1}}\right) 
\nonumber \\
&& - \arctan\left(\frac{1}{\sqrt{\frac{E^{2}r_{o}^{2}}{d^{2}}-1}}\right)\biggl]\Biggr\}
\,\, .
\label{dragging}
\end{eqnarray}
%\end{widetext}
%
This result is the same obtained in \cite{Pineault}, since $q=0$, $\Lambda_{4D}=0$, $E = 1$, and $r_{o}\rightarrow\infty$ ($\Delta\varphi=2Ma/d^2$). In a simple example, when $\Lambda_{4D} = 0$, the metric (\ref{Metrica_Boyer-Lindquist}) is asymptotically flat. Then we can assume $E = 1$ and $r_{o}\rightarrow\infty$, following Pineault \textit{et al.} Thus the dragging, in this example, is 
\begin{equation}
\Delta\varphi  = \int_{\varphi_{i}}^{\varphi_{f}}d\varphi  =  \int_{d}^{\infty}\frac{\left(2Mr-q\right)a}{r^{3}\sqrt{r^{2}-d^{2}}}\ dr
= \left(\frac{2M}{d^{2}}-\frac{\pi q}{4d^{3}}\right)a \,\, , 
\label{Inegral_1}
\end{equation}
and the variation between $a^{(\mu)}$ and $a_{+}^{(\mu)}$ along the congruence is
\begin{equation}
\Delta\chi = - \left(\frac{2M}{d^{2}}-\frac{3\pi q}{8d^{3}}\right)a. 
\label{Delta_chii}
\end{equation}
Consequently the variation of the polarization vector is, according to Eq.~(\ref{Delta}), in the first case, where observer and source are at a distance $d$ from symmetry axis, 
\begin{equation}
\Delta\omega = \Delta\chi+\Delta\varphi=\frac{\pi aq}{8d^{3}}.
\label{Delta_total1}
\end{equation}
When $q=0$, the bulk's influence is absent and we have the same result
in Pineault-Roeder's work \cite{Pineault}. But in the brane world scenario, the existence
of the tidal charge $q$ implies the amplification of the gravitational
effects when $q<0$. In this case, the polarization vector is counter-rotating
with respect to the spin of the black hole.

\subsection{Source and observer in the symmetry axis}

In the second case (see Fig.~\ref{scenarios}, bottom panel), when the source and the observer are in the symmetry
axis, $d=0$ and we have $\Delta\chi=0$ according to Eq.~(\ref{D_xi2}). Then, Eq.~(\ref{dragging})
turns out to be
\begin{gather}
\Delta\omega =\Delta\varphi = \int_{\varphi_{i}}^{\varphi_{f}}d\varphi = \int_{r_{s}}^{r_{o}}\left[\frac{2Mr-q}{r^{4}}+\frac{\Lambda_{4D}}{3}\right]adr \nonumber \\
= a \, \left[M\left(\frac{1}{r_{s}^{2}}-\frac{1}{r_{o}^{2}}\right)-\frac{q}{3}\left(\frac{1}{r_{s}^{3}}-\frac{1}{r_{o}^{3}}\right)-\frac{\Lambda_{4D}}{3}\left(r_{s}-r_{o}\right)\right] \,\, .
\label{dragging_2}
\end{gather}
Again, setting $q=0$ and $\Lambda_{4D}=0$, with $r_{o}\rightarrow\infty$, we recover the Pineault-Roeder and Godfrey \cite{Godfrey} results. In the more general case treated in this work, we observe that the gravitational effects are stronger when $q<0$ and $\Lambda_{4D}>0$.

\section{Final remarks}
\label{remarks}

We have obtained vacuum solutions of the induced gravitational field
equations, deduced by Shiromizu \textit{et al.} \cite{Shiromizu},
with axial symmetry in a Randall-Sundrum brane world type scenario. In the present work, we have used the Kerr-Schild-(A)-dS ansatz to generate axially symmetric solutions on the brane with cosmological constant. Asymptotically de Sitter and anti-de Sitter spacetimes were constructed.

An important result in the solutions obtained is the presence of a tidal charge $q$. This real parameter can be positive or negative. For negative values of the tidal charge, the effects of the bulk on the brane are amplified. The characteristics of the Killing horizons are different compared to the four-dimensional Einsteinian gravity counterparts. The possibility of the negative
values of $q$ provides a mechanism which amplifies the gravitational
effects on the brane. 

In order to analyze possible effects of the extra dimensions in terms of measurable quantities, we have focused on geometrical optics.
With these metrics we studied optical features of these geometries.
In the geometrical optics regime, we used the Pineault-Roeder
approach \cite{Pineault} to characterize the rotation of the polarization vector in the null congruence direction. Again, negative values of $q$ amplify the results. Our work suggests that the analysis of the optical features of rotating black hole candidates might reveal a possible influence of extra dimensions in observable quantities.

Another interesting issue in this context is the study of gravitational radiation from astrophysical sources. According to the present work, a negative tidal charge enhances the value of certain physical observables. It is an open question, for instance, what is the influence of this negative charge is on the frequency and amplitude of gravitational waves for high values of rotation ($a \approx M$). In this regime, a strong gravitational wave emission from highly spinning black holes in compact binaries is expected. Signatures of such events might be detected by the current and upcoming gravitational wave observatories \cite{Brown,Sathyaprakash,Amaro-Seoane}.

\begin{acknowledgments}
This work was partially supported by Conselho Nacional de Desenvolvimento Cient\'{\i}fico e Tecnol\'{o}gico (CNPq), Coordena\c{c}\~{a}o de Aperfei\c{c}oamento de Pessoal de N\'{\i}vel Superior (CAPES), and Funda\c{c}\~{a}o de Amparo \`{a} Pesquisa do Estado de S\~{a}o Paulo (FAPESP), Brazil. 
\end{acknowledgments}

\appendix

\section{Geodesics}
\label{geodesics}

In 1968, Carter \cite{Carter2} showed the separability of the Hamilton-Jacobi
equation for the Kerr-Newman metric and constructed its geodesics
equations. The geodesics equations associated with the metric in Eq.~(\ref{Metrica_Boyer-Lindquist}), with the $(t,r,\theta,\phi)$ coordinate system, can be obtained by the same method used by Carter:
\begin{eqnarray}
\Sigma\dot{t} & = & \frac{(r^{2}+a^{2})P}{\Delta_{r}}-\frac{a}{\Delta_{\theta}}\left[aE\ \sin^{2}\theta-\left(1+\frac{\Lambda_{4D}}{3}a^{2}\right)\Phi\right] \,\, ,
\nonumber \\
\label{t_pt} \\
\Sigma\dot{r} & = & \sqrt{\mathcal{R}} \,\, ,
\label{r_pt} \\
\Sigma\dot{\theta} & = & \sqrt{\Theta} \,\, ,
\label{theta_pt} \\
\Sigma\dot{\varphi} & = & \frac{aP}{\Delta_{r}}-\frac{1}{\Delta_{\theta}}\left[aE-\left(1+\frac{\Lambda_{4D}}{3}a^{2}\right)\textrm{cosec}^{2}\theta\ \Phi\right] \,\, ,
\label{varphi_pt}
\end{eqnarray}
where the dot means an ordinary derivative with respect to the affine
parameter $\mu$, and the functions $P$, $\mathcal{R}$ and $\Theta$ are
\begin{eqnarray}
P & = & \left(r^{2}+a^{2}\right)E - \left(1+\frac{\Lambda_{4D}}{3}a^{2}\right)a\Phi \,\, ,
\nonumber \\
\mathcal{R} & = & P^{2}-\Delta_{r}\left(\pm\delta^{2}r^{2}+K\right) \,\, ,
\nonumber \\
\Theta & = & Q - \cos^{2}\theta \biggl[ a^{2}\left(\pm\Delta_{\theta}\delta^{2}-E^{2}\right) 
\nonumber \\
& & + \left(1+\frac{\Lambda_{4D}}{3}a^{2}\right)^{2}\textrm{cosec}^{2}\theta\ \Phi^{2} \biggl] \,\, .
\label{fun=0000E7=0000F5es_geodesicas}
\end{eqnarray}
The parameter $\delta$ represents the mass of the particle (in the
photon case $\delta=0$), and the signal
plus denotes dS and minus denotes AdS. The $Q$ constant is related to the
Carter's constant $K$. In the present work
\begin{equation}
Q = \Delta_{\theta}K-\left[\left(1+\frac{\Lambda_{4D}}{3}a^{2}\right)\Phi-aE\right]^{2} \,\, .
\end{equation}

In equatorial orbits $Q$ vanishes. The constants $E$ and $\Phi$ are
constants of motion derived by the two Killing vector fields $\xi_{t}$ and $\xi_{\varphi}$ in the geometry with axial symmetry (\ref{Metrica_Boyer-Lindquist}).
That is
\begin{equation}
p_{t}=-E\ \ \ \ \mbox{and} \ \ \ \ p_{\varphi} = \Phi \,\, .
\end{equation}

\section{Quantities in the locally nonrotating frame}
\label{non-rotating}

All important quantities are indicated by parenthesis around the Greek
indices in the LNRF. The components of $k^{\mu}=dx^{\mu}/d\mu=\dot{x}^{\mu}$,
which is tangent to the null congruence, and its projections,  $k^{(\mu)}$, on the LNRF are 
\begin{eqnarray}
k^{(0)} = k^{(t)} & = & e^{\nu}k^{0}=e^{\nu}\dot{t} \,\, , \nonumber \\
k^{(1)} = k^{(r)} & = & e^{\lambda}k^{1}=e^{\lambda}\dot{r} \,\, , \nonumber \\
k^{(2)} = k^{(\theta)} & = & e^{\mu}k^{2}=e^{\mu}\dot{\theta} \,\, , \nonumber \\
k^{(3)} = k^{(\varphi)} & = & e^{\psi}\left(k^{3}-\Omega k^{1}\right) = e^{\psi} \left(\dot{\varphi} - \Omega\dot{t}\right) \,\, .
\label{k_projetado}
\end{eqnarray}
The functions $\nu,\lambda,\mu$, and $\psi$ are listed in (\ref{functions}).

The nonzero components of the connection projected on the LNRF \cite{Bardeen} are
\begin{eqnarray}
\Gamma_{\ \ \ (r)(t)}^{(t)} & = & \Gamma_{\ \ \ (t)(t)}^{(r)}=\partial_{r}\nu e^{-\lambda} \,\, , \nonumber \\
\Gamma_{\ \ \ (\theta)(t)}^{(t)} & = & \Gamma_{\ \ \ (t)(t)}^{(\theta)}=\partial_{\theta}\nu e^{-\mu} \,\, , \nonumber \\
\Gamma_{\ \ \ (\theta)(r)}^{(r)} & = & -\Gamma_{\ \ \ (r)(r)}^{(\theta)}=\partial_{\theta}\lambda e^{-\mu} \,\, , \nonumber \\
\Gamma_{\ \ \ (\theta)(\theta)}^{(r)} & = & -\Gamma_{\ \ \ (r)(\theta)}^{(\theta)}=-\partial_{r}\mu e^{-\lambda} \,\, , \nonumber \\
\Gamma_{\ \ \ (\varphi)(\varphi)}^{(r)} & = & -\Gamma_{\ \ \ (r)(\varphi)}^{(\varphi)}=-\partial_{r}\psi e^{-\lambda} \,\, , \nonumber \\
\Gamma_{\ \ \ (\varphi)(\varphi)}^{(\theta)} & = & -\Gamma_{\ \ \ (\theta)(\varphi)}^{(\varphi)}=-\partial_{\theta}\psi e^{-\mu} \,\, , \nonumber
\end{eqnarray}
\begin{gather}
\Gamma_{\ \ \ (r)(\varphi)}^{(t)}=\Gamma_{\ \ \ (t)(\varphi)}^{(r)}=\Gamma_{\ \ \ (\varphi)(r)}^{(t)} = \Gamma_{\ \ \ (\varphi)(t)}^{(r)} \nonumber \\
=-\Gamma_{\ \ \ (t)(r)}^{(\varphi)}=-\Gamma_{\ \ \ (r)(t)}^{(\varphi)}=\frac{1}{2}\partial_{r}\Omega e^{\psi-\nu-\lambda} \,\, , \nonumber
\end{gather}
\begin{gather}
\Gamma_{\ \ \ (\theta)(\varphi)}^{(t)}=\Gamma_{\ \ \ (t)(\varphi)}^{(\theta)}=\Gamma_{\ \ \ (\varphi)(\theta)}^{(t)} = \Gamma_{\ \ \ (\varphi)(t)}^{(\theta)}
\nonumber \\
= -\Gamma_{\ \ \ (t)(\theta)}^{(\varphi)}=-\Gamma_{\ \ \ (\theta)(t)}^{(\varphi)}=\frac{1}{2}\partial_{\theta}\Omega e^{\psi-\nu-\mu} \,\, , \nonumber \\
\label{Conec=0000E7=0000F5es}
\end{gather}
with $e^{2\nu},e^{\psi},e^{2\lambda}$, and $e^{2\mu}$ given by
\begin{equation*}
e^{2\nu} = \frac{\Sigma\Delta_{r}\Delta_{\theta}}{\Pi\left(1 + \frac{\Lambda_{4D}}{3}a^{2}\right)} \,\, , 
\end{equation*}
\begin{equation*}
e^{2\psi} = \frac{\Xi \sin^{2}\theta}{\Sigma\left(1 + \frac{\Lambda_{4D}}{3}a^{2}\right)} \,\, ,
\end{equation*}
\begin{equation*}
e^{2\lambda} = \frac{\Sigma}{\Delta_{r}} \,\, ,
\end{equation*}
\begin{equation}
 e^{2\mu} = \frac{\Sigma}{\Delta_{\theta}} \,\, ,
\label{functions}
\end{equation}
where
\begin{eqnarray}
\Pi & = & \frac{(r^{2} + a^{2})^{2} \Delta_{\theta} - a^{2} \Delta_{r} \sin^{2} \theta}{\left( 1 + \frac{\Lambda_{4D}}{3}a^{2}\right)} \,\, ,
\label{Pi} \\
\Omega & = &  \frac{a\left[(r^{2} + a^{2}) \Delta_{\theta} - \Delta_{r} \right]}{\Xi} \,\, .
\label{Omega}
\end{eqnarray}

The components of the Riemann tensor in the same frame are
\begin{eqnarray*}
R_{(t)(\varphi)(t)(\varphi)} & = & -R_{(r)(\theta)(r)(\theta)}=Q_{1} \,\, , \\
R_{(t)(\varphi)(r)(\theta)} & = & -Q_{2} \,\, , \\
R_{(t)(r)(t)(r)} & = & -\frac{1}{1-z}\left[Q_{1}\left(2+z\right)-\frac{q}{\Sigma^{2}}+\Lambda_{4D}\right] \,\, , \\
R_{(\varphi)(\theta)(\varphi)(\theta)} & = & \frac{1}{1-z}\left[Q_{1}\left(2+z\right)+\frac{q}{\Sigma^{2}}+\Lambda_{4D}\right] \,\, , \\
R_{(t)(r)(t)(\theta)} & = & R_{(\varphi)(r)(\varphi)(\theta)}=SQ_{2} \,\, , \\
R_{(t)(r)(\varphi)(r)} & = & S\left(Q_{1}-\frac{q}{3\Sigma^{2}}+\frac{\Lambda_{4D}}{3}\right) \,\, , \\
R_{(t)(\theta)(\varphi)(\theta)} & = & -S\left(Q_{1}+\frac{q}{3\Sigma^{2}}+\frac{\Lambda_{4D}}{3}\right) \,\, ,
\end{eqnarray*}

\begin{eqnarray}
R_{(t)(r)(\varphi)(\theta)} & = & -Q_{2}\frac{2+z}{1-z} \,\, ,\nonumber \\
R_{(t)(\theta)(t)(\theta)} & = & \frac{1}{1-z}\left[Q_{1}(1+2z)+\left(\frac{q}{\Sigma^{2}}+\Lambda_{4D}\right)z\right] \,\, , \nonumber \\
R_{(\varphi)(r)(\varphi)(r)} & = & -\frac{1}{1-z}\left[Q_{1}(1+2z)-\left(\frac{q}{\Sigma^{2}}-\Lambda_{4D}\right)z\right] \,\, , \nonumber \\
R_{(t)(\theta)(\varphi)(r)} & = & -Q_{2}\frac{1+2z}{1-z} \,\, , 
\label{Riemann}
\end{eqnarray}
with $Q_{1},Q_{2},S$, and $z$ defined as
\begin{eqnarray}
Q_{1} & = & -\frac{\Lambda_{4D}}{3}+\frac{\left[Mr(r^{2}-3a^{2}\cos^{2}\theta)-q(r^{2}-a^{2}\cos^{2}\theta)\right]}{\Sigma^{3}} \,\, , \nonumber \\
Q_{2} & = & -\frac{\Lambda_{4D}}{3}+\frac{\left[Mr(r^{2}-3a^{2}\cos^{2}\theta)-q(r^{2}-a^{2}\cos^{2}\theta)\right]}{\Sigma^{3}} \,\, , \nonumber \\
S & = & \frac{3a(r^{2}+a^{2})(\Delta_{r}\Delta_{\theta})^{1/2}\sin\ \theta}{\left(1+\frac{\Lambda_{4D}}{3}a^{2}\right)\Pi} \,\, , \nonumber \\
z & = & \frac{\Delta_{r}a^{2}\sin^{2}\theta}{\left(r^{2}+a^{2}\right)^{2}\Delta_{\theta}} \,\, .
\label{Q_S_z}
\end{eqnarray}

\end{document}